\documentclass[aps,pra,twocolumn,noshowpacs,superscriptaddress,floatfix]{revtex4}

\usepackage{amsmath}
\usepackage{amssymb}
\usepackage{multirow}
\usepackage{graphicx}
\usepackage{epsfig}
\usepackage{times}
\usepackage{pdfpages}
\usepackage{url}
\usepackage[colorlinks,citecolor=blue,linkcolor=blue,urlcolor=blue,breaklinks=True]{hyperref}
\usepackage{array}

\newcommand\ket[1]{\lvert#1\rangle}
\newcommand\bra[1]{\langle#1\rvert}

\begin{document}
\title{Reconstruction of the Real Quantum Channel via Convex Optimization}

\author{Xuan-Lun Huang}
\affiliation{State Key Laboratory of Advanced Optical Communication Systems and Networks, School of Physics and Astronomy, Shanghai Jiao Tong University, Shanghai 200240, China}
\affiliation{Synergetic Innovation Center of Quantum Information and Quantum Physics, University of Science and Technology of China, Hefei, Anhui 230026, China}

\author{Jun Gao}
\affiliation{State Key Laboratory of Advanced Optical Communication Systems and Networks, School of Physics and Astronomy, Shanghai Jiao Tong University, Shanghai 200240, China}
\affiliation{Synergetic Innovation Center of Quantum Information and Quantum Physics, University of Science and Technology of China, Hefei, Anhui 230026, China}

\author{Zhi-Qiang Jiao}
\affiliation{State Key Laboratory of Advanced Optical Communication Systems and Networks, School of Physics and Astronomy, Shanghai Jiao Tong University, Shanghai 200240, China}
\affiliation{Synergetic Innovation Center of Quantum Information and Quantum Physics, University of Science and Technology of China, Hefei, Anhui 230026, China}

\author{Zeng-Quan Yan}
\affiliation{State Key Laboratory of Advanced Optical Communication Systems and Networks, School of Physics and Astronomy, Shanghai Jiao Tong University, Shanghai 200240, China}
\affiliation{Synergetic Innovation Center of Quantum Information and Quantum Physics, University of Science and Technology of China, Hefei, Anhui 230026, China}

\author{Ling Ji}
\affiliation{State Key Laboratory of Advanced Optical Communication Systems and Networks, School of Physics and Astronomy, Shanghai Jiao Tong University, Shanghai 200240, China}
\affiliation{Synergetic Innovation Center of Quantum Information and Quantum Physics, University of Science and Technology of China, Hefei, Anhui 230026, China}

\author{Xian-Min Jin}
\thanks{xianmin.jin@sjtu.edu.cn}
\affiliation{State Key Laboratory of Advanced Optical Communication Systems and Networks, School of Physics and Astronomy, Shanghai Jiao Tong University, Shanghai 200240, China}
\affiliation{Synergetic Innovation Center of Quantum Information and Quantum Physics, University of Science and Technology of China, Hefei, Anhui 230026, China}
\date{\today}

\begin{abstract}
Quantum process tomography is often used to completely characterize an unknown quantum process. However, it may lead to an unphysical process matrix, which will cause the loss of information respect to the tomography result. Convex optimization, widely used in machine learning, is able to generate a global optimal model that best fits the raw data while keeping the process tomography in a legitimate region. Only by correctly revealing the original action of the process can we seek deeper into its properties like its phase transition and its Hamiltonian. Thus, we reconstruct the real quantum channel using convex optimization from our experimental result obtained in free-space seawater. In addition, we also put forward a criteria, state deviation, to evaluate how well the reconstructed process fits the tomography result. We believe that the crossover between quantum process tomography and convex optimization may help us move forward to machine learning of quantum channels.
\end{abstract}

\maketitle

Quantum technologies have obtained great advances in recent years, including quantum computation \cite{nielsen2010, koklqc, tangqw} and quantum simulation \cite{qsimulator}, as well as quantum communication \cite{jin,fbqkd,jiling}, where a primary task is to give a mathematical characterization of the physical system. Generally, there are two difficulties for this mission, that is, decoherence of quantum states and loss of information in states manipulation. In order to delve deeper into the evolution mechanism of the system, a reveal of the real quantum process is of paramount importance. In the realm of quantum information science \cite{nielsen2010}, such kind of quantum system characterization is commonly known as quantum tomography, which comprises quantum state tomography (QST) \cite{mofqubit} and quantum process tomography (QPT) \cite{qptresource}. QST is an indispensable method in QPT and the information that lies in the process can be transformed into a mathematical mapping, that is the process matrix, between the input and output sides in QPT. These techniques are widely used in process reconstruction and Hamiltonian estimation \cite{qptcd, qptsolid, qptcnot, qptjose}.

A systematical resource analysis \cite{qptresource} is investigated using different kinds of tomography ways, like standard quantum process tomography, ancilla-assisted quantum process tomography, etc. Through these methods, we can rebuild the quantum process of systems for different quantum tasks, for example, quantum communication channels, giving it a complete characterization.

\begin{figure}
\centering
\includegraphics[width=1\columnwidth]{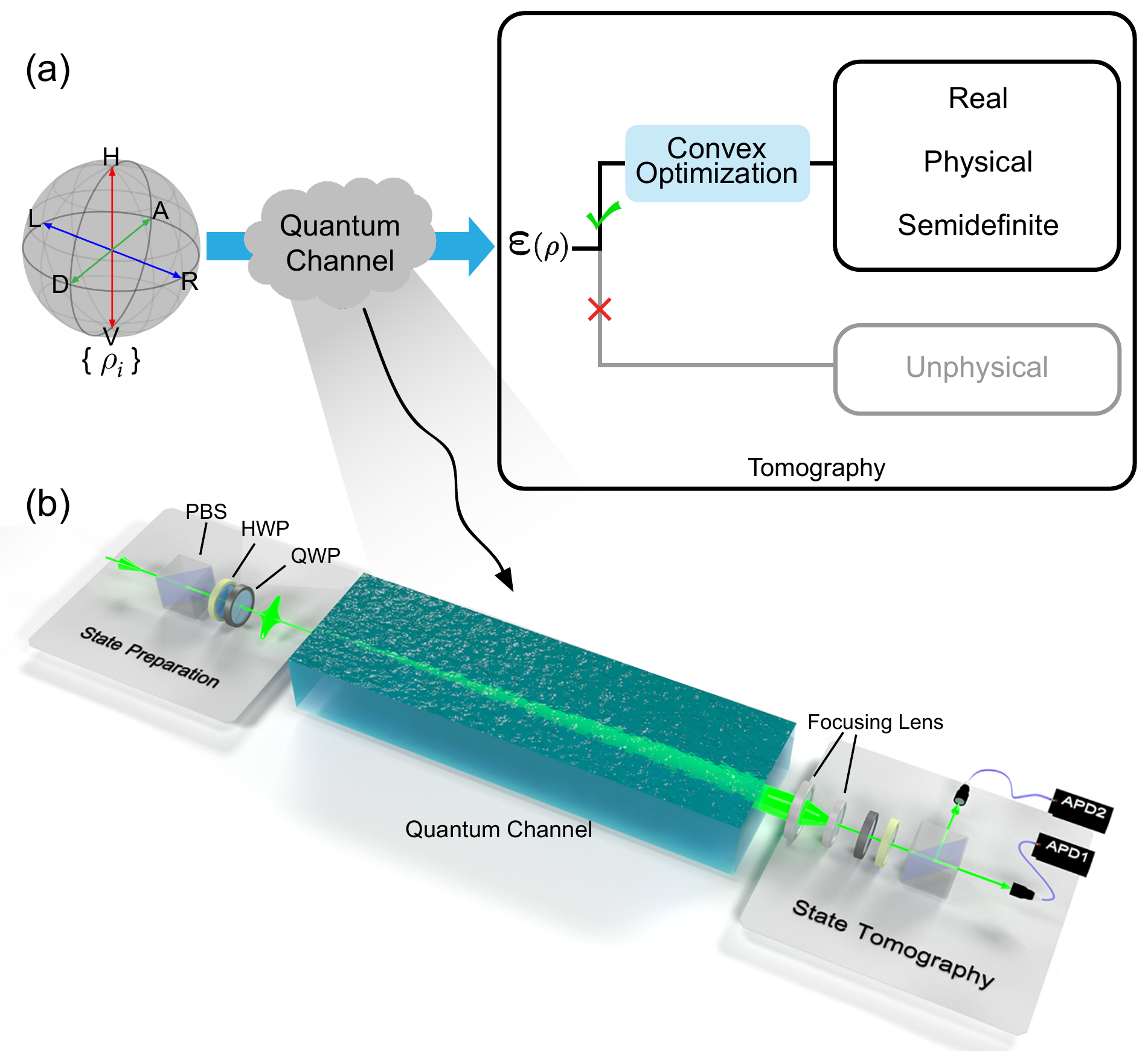}
\label{fig1}
\caption{\textbf{Reconstruction of the real quantum channel.} \textbf{(a)} An unknown quantum channel can be characterized by applying QPT for different input states $\{\rho_i\}$. With the implementation of convex optimization, we can reveal the real and physical quantum process that best fits the tomography result. \textbf{(b)} Our experimental setup of testing a seawater quantum channel. Different input states $\{\rho_i\}$ are prepared by the polarization beam splitter (PBS), the half-wave plate (HWP) and quarter-wave plate (QWP). These states is guided to the quantum channel of seawater and the output is reconstructed by QST.}
\end{figure}

\begin{figure*}[!ht]
\centering
\includegraphics[width=1.93 \columnwidth]{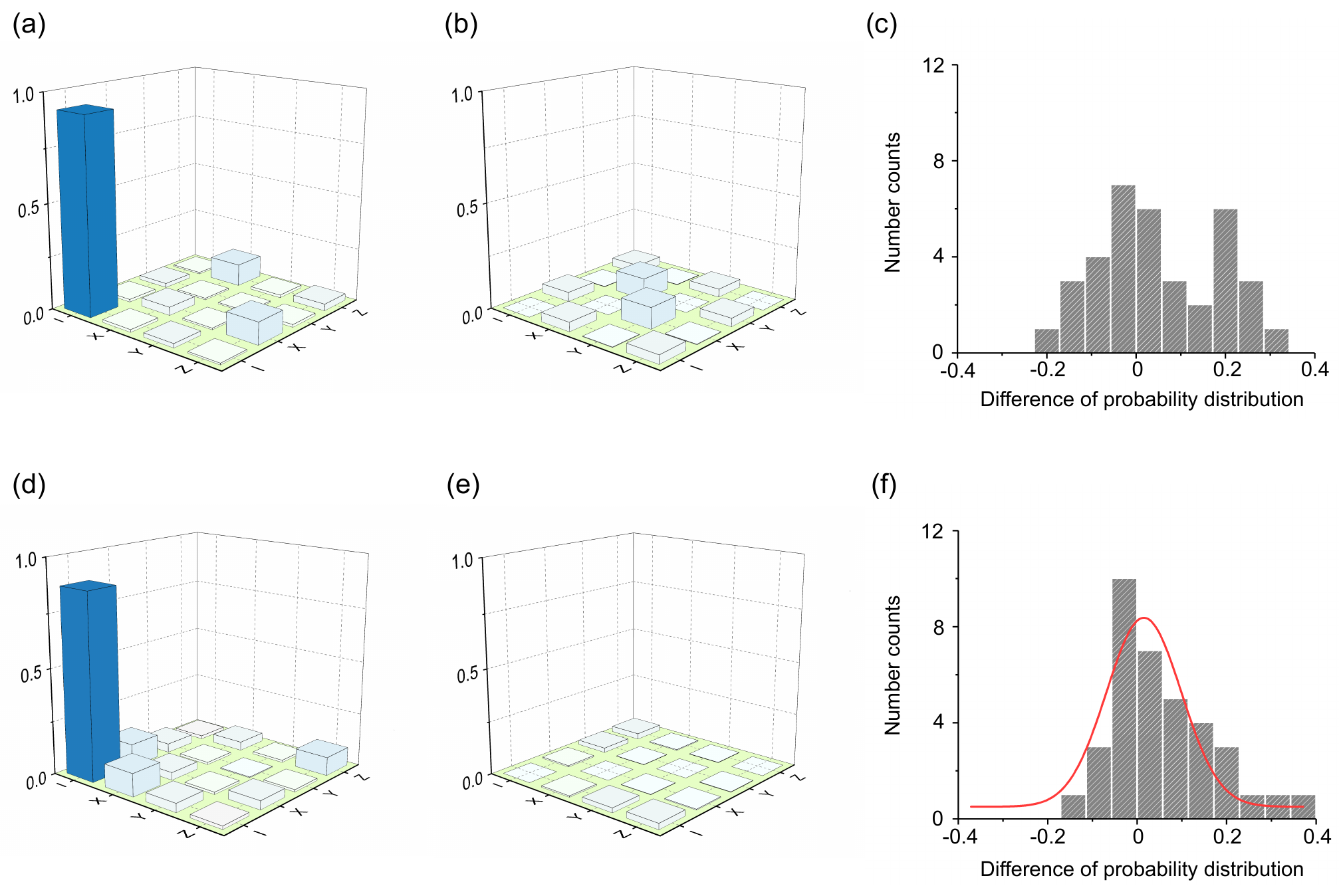}
\label{fig2}
\caption{\textbf{Experimentally reconstructed process matrix of the quantum channel}. \textbf{(a)} and \textbf{(b)} are the real and imaginary parts of the process matrix, reconstructed from standard inversion, which are unphysical. \textbf{(c)} The residual, i.e. the difference between the probability distribution predicted from $\chi^{the}$ and that from the experimental results. \textbf{(d)} and \textbf{(e)} are the corresponding parts of the real process matrix calculated from convex optimization. \textbf{(f)} is the residual of the convex-transformed process matrix $\tilde{\chi}$ which fits a Gaussian distribution better.}
\end{figure*}

Quantum communication, which harnesses the non-cloning theorem, promises unconditionally secure information exchange. Experimental efforts have been made in free-space \cite{jin}, fiber \cite{fbqkd} and underwater \cite{jiling} channels. The key rate and the security rely on the error in channels, which can be caused by decoherence and/or man-made eavesdropping, thus we need to have a full knowledge of the channel. However, standard reconstruction of the channel will lead to an unphysical matrix (Fig. \hyperref[fig1]{1a}), which means the inversion is not the optimal fitting of the raw tomography result. If the reconstruction does not lie in a correct and physical region, it will result in a misjudgement on channels. What's worse, in terms of multipartite communication scheme which requires a participation of many channels, such kind of errors will accumulate exponentially. Even for a standard point-to-point quantum communication channel (Fig. \hyperref[fig1]{1b}), we find that the unphysical issue is unnegligible (Fig. \hyperref[fig2]{2a, 2b} and \hyperref[fig2]{2c}), which forces us to derive an approach to reconstruct real channels in a physical fashion.

We resort to some modern-information-processing technologies, like convex optimization \cite{boyd}, one of kernels in machine learning whose task is to find the parameters in the model that best fits the prior information. The general optimization in machine learning will leads to a local optimal while the convex optimization generates a global optimal. By applying the convex optimization, we will be able to extract the full and correct information from our measurement result. In addition, the combination of convex optimization and quantum information may arouse some interesting applications such as neural-network QST \cite{NNqst} and quantum state classifier \cite{Gao2018}.

\renewcommand\arraystretch{1.5}
\renewcommand{\thetable}{\arabic{table}}
\begin{table*}
\centering
\begin{tabular}{lcccccccc}
\hline\hline
$\quad$ & $\mathrm{I}$ & $\mathrm{II}$ & $\mathrm{III}$ & $\mathrm{IV}$ & $\mathrm{V}$ & $\mathrm{VI}$ & $\mathrm{VII}$ & $\mathrm{VIII}$\\
\hline
$\tilde{F}_{p}$ & $0.934027$ & $0.962625$ & $0.975603$ & $0.997111$ & $0.973731$ & $0.987083$ & $0.973694$ & $0.969428$\\
$\tilde{\Delta}_{avg}$&0.0071&0.0044&0.0011&2.7e-04&9.7e-04&2.4e-04&0.0098&0.0017\\
$\Delta_{avg}^{the}$&0.0113&0.0051&0.0044&6.0e-04&0.0017&6.8e-04&0.0098&0.0033\\
Optimal & $0.0957$ & $0.0451$ & $0.0240$ & $0.005$ & $0.0043$ & $0.0032$ & $0.0753$ & $0.0474$\\
\hline\hline
\end{tabular}
\label{table1}
\caption{\textbf{Measured process fidelities, state deviations and optimal values.} $\tilde{F}_p$ are the fidelities of different seawater samples derived from convex-optimization-based tomography. $\tilde{\Delta}_{avg}$ are the average state deviations (\ref{eq:sd}) of different input states $\{\rho_i\}$, which are obtained from convex optimization, while $\Delta_{avg}^{the}$ are the ones from standard inversion. For a process matrix better fits the real quantum channel, it has a smaller $\Delta_{avg}$. And the last row are the optimal values of the objective function (\ref{eq:qptf}).}
\end{table*}

In this letter, We formulate the reconstruction as a convex optimization problem \cite{boyd,ballocvx} to the real quantum channel, which overcome the unphysical issue in standard QPT. Besides, solving the optimization problem of a convex function leads to the global optimal, which means we can accurately determine the true action of the channels on different input states. After obtaining the process matrix, we introduce a criteria, state deviation, to measure the consistency between the reconstructed and experimental results.

The purpose of QPT \cite{nielsen2010,qptresource} is to determine the unknown process $\mathcal{E}$, which can be expressed in the operator-sum representation, such that, $\mathcal{E}(\rho)=\sum_{i}E_i \rho E_{i}^{\dag}$, where $\{E_{i}\}$ are the operator elements of the process operation $\mathcal{E}$. During the procedure of QPT, it needs to prepare the quantum system in a set of quantum states and subject them respectively to the channel. After that, QST is performed to measure the output states and then the operation $\mathcal{E}$ is fully characterized by a linear extension of these states. The density matrixes of the input states and the measurement projectors must each form a basis set of the state space, and $d^2=2^{2n}$ \cite{mofqubit} elements are required in each set supposing the state space has $d$ dimensions. For our one-qubit seawater channel \cite{jiling}, 4 elements $\{\rho_{i}\}$, namely, $\{\rho_{H}, \rho_{V}, \rho_{D}, \rho_{R}\}$, are chosen as the input and output states. Through this paper, 4 polarization states $\{H,V,D,R\}$ denote the qubit $\{\ket{1},\ket{0},(\ket{1}+\ket{0})/\sqrt{2},(\ket{1}+i\ket{0})/\sqrt{2}\}$, respectively.

In practice, experimental results involve numerical data rather than operators, so it is convenient to describe the process using a transformation matrix. There exists two kinds of expressing methods, the Choi matrix \cite{ballocvx} and the $\chi$ matrix \cite{qptresource}. Though it is convertible between these two schemes \cite{2qubit}, the $\chi$ matrix is much more preferable, same in this work, due to its straightforward representation. Hence, the operator-sum representation can be extended as
\begin{equation}
\mathcal{E}(\rho)=\sum_{mn}\chi_{mn}\tilde{E}_{m} \rho \tilde{E}_{n}^{\dag}  \label{eq:chisum}
\end{equation}
where $\{\tilde{E}_{i}\}$, serving as a fixed set of operators on the Hilbert space, expands the operator elements as $E_{i}=\sum_{m}e_{im}\tilde{E}_{m}$, and $\chi_{mn}=\sum_{i}e_{im}e_{in}^{*}$ are the entries of the $\chi$ matrix, which is positive Hermitian, uniquely describing the process $\mathcal{E}$. And $\{\tilde{E}_{i}\}$ is chosen to be the Pauli basis, i.e. $\{I,\sigma_{x},\sigma_{y},\sigma_{z}\}$.

However, the output density matrix and the process matrix calculated from the tomographic data generally violate the condition that, for a physical system, the quantum process is supposed to be completely positive and non-trace-increasing, so that more constraints should be added to the reconstruction process.

Considering the QST, the output state should be a density matrix in the output Hilbert space. This means it has trace equal to one and be semidefinite \cite{nielsen2010}. Suppose the noise on the measurement has a Gaussian probability \cite{mofqubit}, then the real state can be reconstructed by minimizing a convex function:
\begin{equation}
\mathcal{L}=\sum_{i=1}^{d^2}\frac{[N\bra{\psi_{i}}\rho\ket{\psi_{i}}-n_i]^2}{2N\bra{\psi_{i}}\rho\ket{\psi_{i}}},  \label{eq:qstf}
\end{equation}
where $N$ is the number of total received photons, $\ket{\psi_{i}}$ is the projecting state and $n_i$ is the corresponding count. By solving the convex optimization problem shown below, we can reconstruct the real state $\{\tilde{\rho_i}\}$ from the output side:
\begin{equation}\label{eq:qstcvx}
\begin{split}
\min_{\rho} &\quad \mathcal{L}(\rho)\\
s.t. &\quad \mathrm{Tr}(\rho)=1\\
&\quad \rho \succeq 0
\end{split}
\end{equation}

As for QPT, on the one hand, $\chi$ is positive and Hermitian. On the other hand, $\chi$ represents a trace-preserving process, namely $\mathrm{Tr}(\sum_{mn}\chi_{mn}\tilde{E}_{m} \rho \tilde{E}_{n}^{\dag})=1$, which means $\sum_{mn}\chi_{mn}\tilde{E}_{n}^{\dag}\tilde{E}_{m}=I$. For the operator basis chosen in this paper, it can be concluded that $\mathrm{Tr}(\chi)=1$. Then the tomographic results can be used to reveal the real quantum channel by finding a matrix $\tilde{\chi}$ that matches the theoretical probability with the experimental distribution best. Actually, the best fit $\tilde{\chi}$ is achieved by minimizing a least square objective function \cite{qptcnot}:
\begin{equation}\label{eq:qptf}
\begin{split}
\mathcal{F}(\chi) &= \sum_{i,j=1}^{d^2}\frac{1}{N^2} \Big[ n_{ij}-N\sum_{m,n=0}^{d^2-1}\bra{\psi_j}\tilde{E}_m\ket{\phi_i} \\
&\quad \bra{\phi_i}\tilde{E}_n\ket{\psi_j}\chi_{mn} \Big]^{2},
\end{split}
\end{equation}
where $\ket{\phi_i}$ is the input state and $\ket{\psi_j}$ is the measuring state while $n_{ij}$ is the corresponding photon counts. Combined with the constraints for the process matrix, we can reveal the real quantum channel $\tilde{\chi}$ through this optimization:
\begin{equation}\label{eq:qptcvx}
\begin{split}
\min_{\chi} &\quad \mathcal{F}(\chi)\\
s.t. &\quad \mathrm{Tr}(\chi)=1\\
&\quad \chi \succeq 0
\end{split}
\end{equation}

For an ideal one-qubit channel, $\hat{\chi}$ should have zero entries except for $\hat{\chi}_{00}=1$. As the simple inversion (process tomography scheme in Ref. \cite{nielsen2010} on page 393) will lead to an unphysical matrix. We then use the convex optimization to reconstruct the process matrix. The obtained $\tilde{\chi}$ can be used to evaluate the reconstruction matrix relative to the ideal one by deriving the process fidelity $F_p=\mathrm{Tr}(\sqrt{\sqrt{\tilde{\chi}}\hat{\chi}\sqrt{\tilde{\chi}}})$. Fidelities of different seawater samples \cite{jiling} are shown in Table \hyperref[table1]{1}. Applying the simple inversion method \cite{nielsen2010}, we can get the theoretical process matrix $\chi_{the}$ of sample I, whose absolute real and image values are shown in Fig. \hyperref[fig2]{2a} and \hyperref[fig2]{2b}.

However, when we calculate the eigenvalues of this process matrix, we find that the eigenvalues are $\{0.9263, 0.1941, -0.0203, -0.1001\}$. Process matrix of a physical system is supposed to be positive semidefiniteness, which implies that all of the eigenvalues must be positive and real. Obviously, from the above negative eigenvalues, we can see that the process matrix violates the physical condition. While using the convex optimization method with positive and $\mathrm{Tr}(\chi)=1$ constraints, we can get a physical process matrix, whose eigenvalues are $\{0.1098, 0.8902, 1.3033\mathrm{e}\!-\!14, 1.8719\mathrm{e}\!-\!14\}$, indicating that convex reconstruction gives a legitimate process matrix. Its Pauli basis representation of the real and image parts are shown in Fig. \hyperref[fig2]{2d} and \hyperref[fig2]{2e}, respectively. In addition, we list the minimal value of the objective function (\ref{eq:qptf}) in Table \hyperref[table1]{1}. The time to find the optimal (=0.0957) of the seawater quantum channel result is only about 2 seconds in a dual-core laptop.

\begin{figure}
\centering
\includegraphics[width=1\columnwidth]{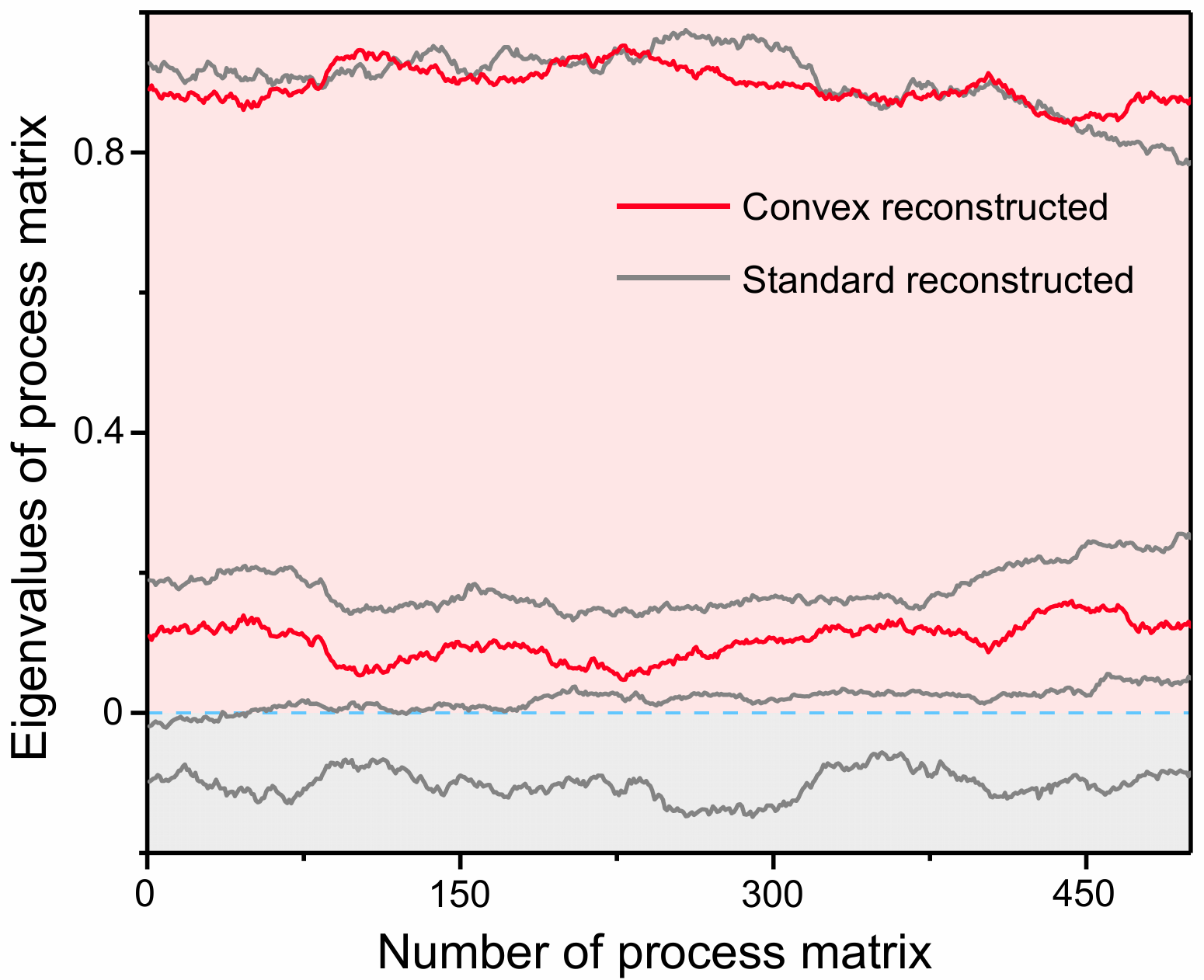}
\label{fig3}
\caption{\textbf{Eigenvalues of the process matrix.} The red lines are eigenvalues of the convex reconstructed process matrix, which are all positive. The gray lines are the eigenvalues of the reconstructed process matrix with the standard inversion, some of which are negative. For a physical matrix, all of its eigenvalues should be in the legitimate region (pink area).}
\end{figure}

Moreover, we can use the Monte Carlo Method to put an error measure for the quantum process. In a standard QPT of a one-qubit channel, it needs $16$ counts number $\{n_1, n_2,\dots, n_{16}\}$ to reconstruct the process matrix. Assume the counts are subject to Poisson distribution, we can obtain a new photon counts $\{n_1\pm\sqrt{n_1}, n_2\pm\sqrt{n_2},\dots, n_{16}\pm\sqrt{n_{16}}\}$. And then we can get different process matrix by randomly choosing the counts from the Poisson distribution. Using this method, we get $F_p=0.9340\pm0.0178$. In Fig. \hyperref[fig3]{3}, we draw the eigenvalues of the two process matrixes, which can be served as a measure for the decoherence \cite{controldeco}. Because the eigenvalues fluctuate in a small amount, it indicates that the quantum channel barely costs decoherence to the qubit. As we can see from the all positive eigenvalues of the convex-optimization-reconstructed process (red lines) in Fig. \hyperref[fig3]{3}, our convex-based QPT keeps the reconstruction in a legitimate region (pink area), while some eigenvalues of the traditional reconstruction (gray lines) lie in the unphysical region (gray area).

Then we examine the residual \cite{qptcnot} (Fig.\,\hyperref[fig2]{2f}), which is the difference between the probability distribution of 6$\times$6 results predicted from $\tilde{\chi}$ and that from the experimental result. The 6$\times$6 results are obtained by preparing 6 input states $\{H,V,D,A,R,L\}$, i.e the horizontal, vertical, diagonal, antidiagonal, right-circular and left-circular polarization qubit, and projecting them to these 6 states respectively. We can find that the distribution fits a Gaussian function with $\sigma=0.16749 \pm 0.03018$, which is consistent with the assumption that the noise on the measurement has a Gaussian probability, while that calculated from the standard inversion (Fig 2c) does not fit a Gaussian distribution. The instance that the residual is over 0.4 in Fig. \hyperref[fig2]{2f} may arise from the fact that the input state is not ideal.

Typically, common optimization problem might fall into a local optimal, therefore, we loose the constraint to be $\mathrm{Tr}(\chi)\leq 1$, a convex function, making the whole optimization a convex problem \cite{boyd}, whose local optimal is the global optimal. And then we also get the same $\tilde{\chi}$, meaning that this is the best fit of the tomographic result. In addition, we use the \emph{minimize} package (with method conjugate gradient) of Scipy \cite{scipy} in python to minimize the objective function (\ref{eq:qptf}) by parameterizing the $\chi$ matrix in a 16 elements vector $\vec{t}$ \cite{mofqubit} and get a similar result while in a much longer time than our convex optimization method.

Further, we test the performance of the convex optimization by comparing the output states predicted by $\tilde{\chi}$ with the experimentally determined and simple inversion calculated states for the inputs $\{\rho_i\}$. Here, state fidelity is insufficient to determine how well the process matrix fits the experimental results because the first element of the density matrix will dominate the trace. We introduce a more appropriate metric, average state deviation $\Delta_{avg}$ to characterize the process. The state deviation is defined that
\begin{equation}\label{eq:sd}
\Delta=\sum_{ij}^{d^2}\frac{(\tilde{\rho}_{ij}-\rho_{ij}^{e})^2}{d^2},
\end{equation}
where $\{\tilde{\rho}_{ij}\}$ are the elements of the predicted density matrix while $\rho_{ij}^{e}$ are the experiment-determined. And then $\Delta_{avg}$ can be obtained by averaging all the input states $\{\rho_i\}$. For a process matrix that fits the raw data better, it should have a smaller state deviation. And the state deviations of different seawater samples are listed in Table \hyperref[table1]{1}. By comparing with the ones calculated from the standard inversion, the process matrix reconstructed from our method reveals the quantum channel better.

In summary, we reconstruct the real quantum channel of seawater including the output states and the process matrix by using convex optimization methods. We introduce the state deviation and show that the results of convex optimization fit the tomography result better and well stay in a physical region. Because the elements in the process matrix represent different decoherence processes, like $\sigma_x\rho\sigma_y$ or $\sigma_x\rho\sigma_y$, we manage to obtain the correct and reliable information of the channel through these convex-optimization-based reconstruction, which may be used to evaluate the decoherence and Hamiltonian \cite{1qubit, 2qubit, pmHi}. These characteristics distinguish different channels from each other those can serve as the ``identity labels" in machine learning. With this fundamental preparation the combination of quantum tomography and convex optimization may trigger more new applications, representing a step further to quantum machine learning \cite{NNqst,Gao2018,qml}.

\subsection*{Acknowledgments}
The authors thank Jian-Wei Pan for helpful discussions. This work was supported by National Key R\&D Program of China (2017YFA0303700); National Natural Science Foundation of China (NSFC) (11374211, 61734005, 11690033).


\end{document}